\documentclass[11pt]{article}
\usepackage[T1]{fontenc}
\usepackage[utf8]{inputenc}
\usepackage{lmodern}
\usepackage[margin=1in,includehead]{geometry}
\usepackage{graphicx}
\usepackage{amsmath,amssymb,amsfonts}
\usepackage{booktabs}
\usepackage{array}
\usepackage{tabularx}
\usepackage[natbibapa]{apacite}
\usepackage{float}
\newcommand{\uptheta}{\theta}
\newcommand{\DagFont}{\rmfamily}
\usepackage{tikz}
\usetikzlibrary{fit,positioning,arrows.meta}
\usepackage[font=footnotesize,labelfont=bf,width=0.9\textwidth]{caption}
\usepackage{subcaption}
\usepackage[hidelinks]{hyperref}
\newcommand{\cref}[1]{\autoref{#1}}
\newlength{\DAGNodeDiam}
\newlength{\DAGHsep}
\newlength{\DAGVsep}
\newlength{\DAGHalfHsep}

\tikzset{
  dag node/.style={
    circle, draw,
    line width=0.4pt,
    minimum size=\DAGNodeDiam,
    inner sep=0pt,
    align=center,
    font=\DagFont\fontsize{7pt}{8pt}\selectfont
  },
  dag edge/.style={
  -{Stealth[length=2.4mm,width=1.6mm]},
  line width=0.4pt
  }
}

\tikzset{
  dag box/.style={
    draw,
    dashed,
    rounded corners=2pt,
    line width=0.4pt,
    inner sep=3pt
  },
  dag label/.style={
    font=\DagFont\fontsize{6.5pt}{7.5pt}\selectfont,
    align=center
  }
}

\usepackage{authblk}

\title{Turning Old Anomalies Into New Evidence\\[0.3em]
\large Evidential Reconfiguration and Dark Matter in 1974}

\author[1,2,3,*]{Simon Allz\'en}

\affil[1 ] {Department of Philosophy,
Stockholm University}
\affil[2 ]{Institute of Physics,
University of Amsterdam}
\affil[3 ]{Vossius Center for the History of Humanities and Sciences,
University of Amsterdam}
\affil[*]{Correspondence: simon.allzen@philosophy.su.se}

\date{}
\date{}

\begin{document}

\maketitle

\begin{abstract}
\small \noindent The 1974 papers by \cite{ostriker_size_1974} and \cite{einasto_dynamic_1974} are regarded as pivotal in establishing the evidential case for dark matter, although both relied on already existing observational data. By what means did they transform known observational anomalies into evidential support for dark matter? I argue that this transformation arose through \textit{evidential reconfiguration}: extended galactic halos supplied a physical connection between galaxy rotation curves and cluster mass discrepancies, expanding the context of assessment so that results previously investigated locally could constrain one another. The significance of this connection was assessed against differing cosmological commitments and unresolved local dynamical assumptions, which help explain both the emphases of the two papers and their reception. To explicate its contribution to confirmation, I draw on Myrvold's Bayesian account of unification and use a toy model in which taking one result into account changes the assessment of a shared halo parameter, thereby altering expectations about the other. Comparison with an uncoupled model makes the resulting support explicit, while variations of the model show why a shared parameter alone is insufficient: the contribution depends on the connecting assumptions and what competing models predict. The novelty relevant to the old-evidence problem thus lies in what familiar observations became jointly capable of supporting. This reconstruction assesses that support once the connection is specified, while leaving a full account of historical belief change open.
\end{abstract}

\section{Introduction}

\noindent The nature of $\approx 80\%$ of the matter content in the universe remains unknown. Its presence is inferred from gravitational anomalies across a range of cosmic scales, where the observed motions and structures cannot be accounted for by ordinary matter alone. This putative mass component, known as dark matter, is a core posit of the standard model of cosmology, the $\Lambda$CDM (concordance) model.\footnote{$\Lambda$ denotes dark energy, and CDM stands for cold dark matter, where `cold' refers to the negligible free-streaming length of the dark matter component.} Dark matter is invoked to explain the motions of galaxies, the dynamics of galaxy clusters, and the large-scale distribution of matter in the universe. To date, it has eluded non-gravitational detection, leaving its nature radically underdetermined and making it one of the major open problems in contemporary physics. It nevertheless remains deeply embedded in $\Lambda$CDM and its existence is widely accepted by cosmologists.\footnote{This tension makes dark matter a fertile case for philosophers of science. See, e.g., \citet{vanderburgh_dark_2003, vanderburgh_methodological_2005, vanderburgh_quantitative_2014, vanderburgh_interpretive_2014}, \cite{sus2014}, \cite{jacquart_cdm_2021}, \citet{merritt_cosmological_2021, merritt_mond_2021}, \cite{de_baerdemaeker_jump_2020}, \cite{de_baerdemaeker_method-driven_2021}, \cite{de_baerdemaeker_mond_2022}, \citet{antoniou_robustness_2023, antoniou_why_2025}, \cite{duerr_methodological_2023}, \cite{wolf_navigating_2025}, \citet{allzen_scientific_2021, allzen_dark_2026}, \cite{martens_dark_2022}, \cite{martens_integrating_2022}, and \cite{vaynberg2024realism}.} 

Textbook histories suggest that dark matter's central role is the result of a largely linear progression of accumulating evidence. Such accounts can make it appear that each observation was treated as direct evidence for an already established hypothesis. In practice, however, the relevant observations were initially often regarded as isolated, context-specific anomalies, not systematically integrated within a clearly defined framework or working hypothesis. The philosophical question is how these local discrepancies became coherent \textit{evidence} for an ontological posit, and under what conditions such a change is justified. From a Bayesian perspective, this question becomes especially pressing when the observations were already accepted and used in constructing the hypothesis: how could their synthesis strengthen its evidential support without introducing new observational data? This paper provides a Bayesian explanation of that transformation in the 1974 dark matter synthesis.

I build on the historical work by \cite{de_swart_how_2017} and \cite{de_swart_closing_2020}, who identify two seminal papers from 1974 as pivotal for the acceptance and establishment of the dark matter hypothesis. I argue that their proposed halo framework connected phenomena previously treated as local and disconnected, allowing the available observations to jointly constrain a missing-mass hypothesis. I call this transformation \textit{evidential reconfiguration}, and reconstruct its confirmatory contribution by comparing coupled and uncoupled models of the same results. Understanding why scientists pursued the connection also requires attention to their theoretical commitments. For some cosmologists, non-empirical considerations made the cosmic mass budget especially pressing. Their role in motivating the synthesis raises a question about justification alongside the question about confirmation: when were those commitments defensible, and what did the resulting connection give scientists reason to believe?

\subsection{Overview}

\noindent Two features of the reasoning in the period highlighted by \cite{de_swart_how_2017} and \cite{de_swart_closing_2020} structure the two main parts of the paper.

The first concerns \textit{theory assessment}. In the mid--1970s, researchers had access to the same observational results, but differed in how strongly they judged these observations to support a missing-mass hypothesis. For some cosmologists, background commitments, including a preference for a closed or near-critical universe, made the mass-budget question especially consequential. Researchers concentrating on local astronomical problems placed greater weight on unresolved dynamical assumptions. These concerns overlapped across disciplines and could affect both prior assessments and judgments about which cross-domain connections were defensible.\footnote{For accounts of indirect or meta-empirical assessment, see \citet{dawid_string_2013, dawid_significance_2019} and \cite{dawid_no_2015}.} Whether the resulting differences were \textit{rationally permissible} depends on the justification of the commitments involved. To assess a background constraint, I suggest asking whether it:

\begin{itemize}
    \item[(a)] is independently motivated within an established modeling context;
    \item[(b)] has an established history of combining separate phenomena and models into a globally coherent framework that is empirically supported; and
    \item[(c)] is treated as defeasible rather than as dogma.
\end{itemize}

\noindent The second concerns \textit{unification}. The 1974 studies proposed extended galactic halos whose masses and extents also mattered to cluster and cosmological mass accounting. What justified treating this connection as support for missing mass? \citet{myrvold_bayesian_2003, myrvold_evidential_2017} gives us a Bayesian way to answer that question by distinguishing a common-origin explanation from a hypothesis that makes distinct results informative about one another: \textit{the connection changed what the results could jointly constrain.} I specify the probabilistic requirements for this contribution, including the independence assumption used to represent an uncoupled comparison.

\textit{Evidential reconfiguration} names the historical transformation of local anomalies into jointly informative evidence, and \textit{structural novelty} identifies the new connection responsible for it. The discussion of community differences explains the background against which the connection was pursued and assessed. The Bayesian reconstruction specifies its possible contribution to support, allowing us to distinguish that contribution from the further problem of modeling historical changes in credence when we return to old evidence.

\section{The standard historical narrative}

\noindent Textbook and popular narratives of dark matter often begin with Zwicky’s application of the virial theorem to the Coma Cluster in the 1930s, then follow a small number of later breakthroughs presented as progressively establishing the hypothesis. If the cluster was bound and approximately equilibrated, its galaxy velocities required more mass than the luminosities appeared to supply. The inference depended on that dynamical interpretation, not on directly observing a bound state. In retrospect, this is often treated as the earliest observational basis for non-luminous mass.\footnote{At the time, \citet{zwicky_how_1933, zwicky_masses_1937} did use the term \textit{``Dunkle Materie''} for unseen mass. However, its conceptual content bears little resemblance to the current concept of dark matter (see \cite{allzen_dark_2024-1} for a historical overview).}

The narrative then emphasizes the rotation-curve work of \cite{rubin_rotation_1970} in the 1970s: rotational velocities in spiral galaxies did not decline with distance from the galactic center as expected by Keplerian dynamics. Subsequent observations by \cite{roberts_comparison_1973}, \cite{bosma_distribution_1978}, and others strengthened the case that many spiral galaxies exhibit approximately flat rotation curves (See Fig.\ref{fig-rotationcurves}). Stars far from the galactic center orbit nearly as quickly as those closer in, in tension with the decline expected from visible matter alone under simple dynamical assumptions. Textbook presentations often treat these results as decisive vindication of dark matter, although Rubin herself was cautious about any causal interpretation of her measurements \citep{rubin_brief_2004}, and many astronomers initially explored explanations that did not require positing a new exotic mass component.

The story is completed by incorporating later developments such as gravitational lensing, large-scale structure formation, and the Bullet Cluster observation \citep{clowe_direct_2006}. These are presented as extensions of the Zwicky--Rubin sequence, progressively strengthening the hypothesis until it became a central posit of contemporary cosmology. This narrative has an implicitly, and perhaps accidentally, \textit{Bayesian} structure: an early hypothesis is followed by a linear series of results treated as evidence suitable for conditionalization. I do not claim that historical presentations organized in this way are illegitimate. They serve important pedagogical purposes, using hindsight to select episodes that connect to the current concept and explanatory role of dark matter. They nevertheless obscure an epistemically prior question: when, and how, did particular results come to function as evidence for missing mass rather than as local anomalies or modeling artifacts?

The 1974 episode examined by \cite{de_swart_how_2017} and \cite{de_swart_closing_2020} brings this question into focus. The two papers contributed to establishing the dark matter hypothesis without introducing new observational data. My aim is to complement the historical claim \textit{that} the 1974 papers mattered with an epistemic account of \textit{why} they mattered. Because the relevant observations were already accepted, ordinary Bayesian conditionalization on them cannot explain this transformation. We need to examine changes in which assumptions were held fixed, and how connecting existing observations transformed anomalies into evidence.

\section{Constraints and the mass deficit}

\noindent To understand why the dark matter (or missing-mass) hypothesis gained traction in both cosmology and astronomy, it is necessary to reconstruct the cosmological background that shaped theory assessment in the decades leading up to the 1970s. In the early and mid-20th century, cosmology was often guided more by theoretical considerations than by the observation-driven practices of astronomy, and debates frequently appealed to methodological and philosophical commitments \citep{de_swart_closing_2020}. One example is the early acceptance of the cosmological principle, formulated by \cite{milne_world_1932}, which asserts that the universe is homogeneous and isotropic on sufficiently large scales. Today, this principle has substantial empirical support, but in the 1930s it was adopted in part for philosophical reasons. Combining the cosmological principle with general relativity led to relativistic cosmology, which by the 1960s had become the predominant framework, drawing on theoretical work by Alexander Friedmann and Georges Lemaître, and observational work by Edwin Hubble \citep{de_swart_closing_2020}.\footnote{See \cite{bondi_steady-state_1948}, who argued that one cannot assume \textit{a priori} that general relativity applies globally without empirical justification, an objection to the early philosophically motivated cosmological assumptions.} By the 1950s and 1960s, two non-empirical commitments in particular structured cosmological background assumptions: the \textit{cosmological principle} and a \textit{Machian inheritance}, according to which local inertia is determined globally by the total mass of the universe. In relativistic cosmology, the latter was often taken to favor a closed, finite universe.

These commitments encouraged some cosmologists to interpret empirical results relative to models partly favored for non-empirical reasons. These preferences included both spatially closed models ($k=+1$) and models at critical density. In matter-only cosmologies with zero cosmological constant, $\Omega=1$ corresponds to the spatially flat Einstein--de Sitter model, at the boundary sometimes described as ``just closed'' in historical discussions. Keeping this historical usage distinct from positive spatial curvature clarifies the related preferences documented by \cite{de_swart_closing_2020}:

\begin{quote}
Nearing the end of the 1960s, a closed universe was still a much-preferred model for many cosmologists and relativists. “\textit{Philosophically}, there might be a preference”, Wolfgang Rindler wrote in 1967, “the choice k = 1 [a positively curved universe] might appear desirable. It implies closed space sections that would, in some sense, validate Mach’s principle according to which the totality of matter in the universe and nothing else determines the local inertial frames” (Rindler 1967: 29--30, emphasis in original). [$\dots$] The implicit preference for a closed universe was also expressed in observational studies. ``One would particularly like to know whether there is enough mass to close the universe'', Princeton physicists Peebles and Partridge wrote in 1967, in a piece on estimating the mass density of the universe (Peebles and Partridge 1967: 713). \cite[pp.~274--275]{de_swart_closing_2020}
\end{quote}

\noindent These non-empirical commitments mattered as \textit{constraints on theoretical modeling}: they shaped which questions and problems seemed pressing and which theoretical ideas were worth pursuing to resolve them. Weinberg captures this reasoning:

\begin{quote}
 If one tentatively accepts the result that $q_0$ is of order unity [$\Omega \gtrsim 1$], then one is forced to the conclusion that a mass density of about $2\times10^{-29}$~g/cm$^3$ must be found somewhere outside the normal galaxies. But where? \cite[p.~478]{weinberg_gravitation_1972}
\end{quote}

\noindent These commitments were revisable in principle, but taking them seriously made certain problems appear more consequential, as Peebles and Partridge’s ``\textit{whether}'' and Weinberg’s ``\textit{where}'' illustrate. \cite{PeeblesPartridge1967} ask whether there is enough mass to close the universe, while \cite{weinberg_gravitation_1972} presses the follow-up question of where that mass must be found. By the early 1970s, near-critical cosmological models functioned for some authors as a defensible target and for others as a background commitment with independent motivations. Either way, these commitments helped create a context in which missing mass could be pursued as a candidate astronomical solution to a cosmological problem; a problem which was characterized as such entirely without reference to the arrival of new empirical discrepancies or observations.

Against this background, two research groups independently arrived at similar proposals in 1974. \cite{ostriker_size_1974} and \cite{einasto_dynamic_1974} looked to galactic astronomy for a common dynamical account of missing mass on galactic and cluster scales, with implications for the cosmic mass budget. The emphasis of the two papers, however, was not identical: \cite{ostriker_size_1974} foreground questions about cosmological density and closure considerations from the outset, while \cite{einasto_dynamic_1974} foreground the cluster virial mass discrepancy and treat massive coronas as a way to reduce it \citep[310]{einasto_dynamic_1974}. Both groups drew on two kinds of phenomena that were already well known but typically treated as separate research problems in different disciplines. The methods and uncertainties differed, and there was no established view in which the phenomena could be considered naturally connected, let alone bear on the same hypothesis in tandem.

\begin{figure}[H]
    \centering
    \includegraphics[width=0.70\textwidth]{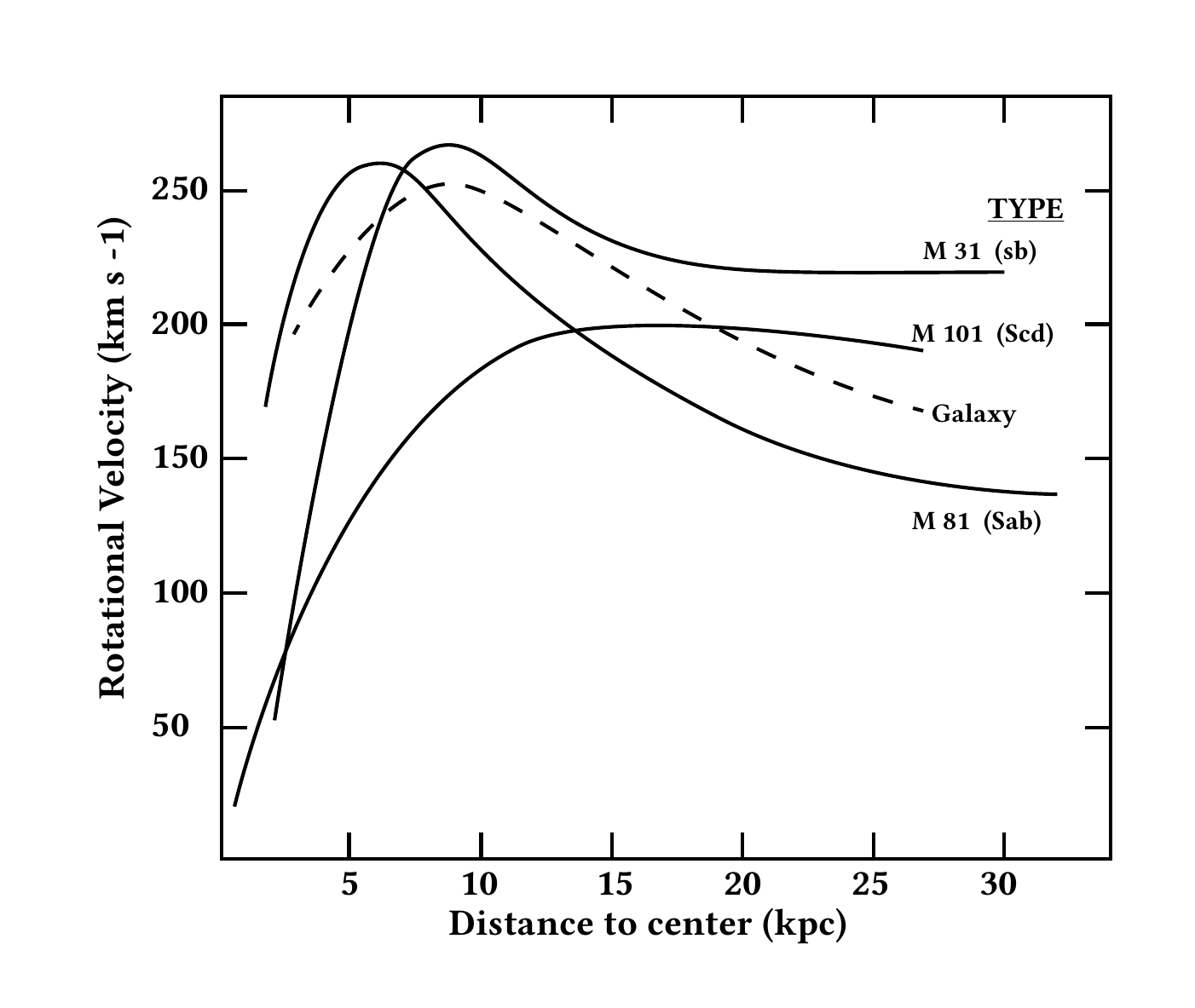}
    \vspace{-0.3cm}
    \caption{Rotation curves of three spiral galaxies (solid lines) as a function of distance from the center. (The Milky Way’s expected Keplerian decline is shown for reference as the dashed line.) The flatness of the curves at large radii indicates the presence of unseen mass.}
    \label{fig-rotationcurves}
\end{figure}

\begin{enumerate}
    \item \textit{Galaxy dynamics}.\\
Radio observations of the 21-cm hydrogen line allowed astronomers to measure rotation speeds of spiral galaxies at radii well beyond the visible stellar disk. By the early 1970s, abundant rotation-curve data were available. Keplerian dynamics suggested that rotation speeds should decrease with radius beyond most of the galaxy’s luminous matter (see \cref{fig-rotationcurves}, reconstructed from the original plot in \cite{roberts_comparison_1973}). Empirically, however, many spiral galaxies showed approximately flat rotation curves. This behavior is \textit{consistent} with additional gravitating mass at large radii, but did not by itself establish any single ontological conclusion. In practice, these results were often treated as local anomalies, whose significance depended on auxiliary assumptions in astronomy about tracer populations, mass-to-light ratios, gas distributions, and dynamical modeling.

   \item \textit{Galaxy cluster dynamics}.\\
The virial mass discrepancy had long been known. \citet{zwicky_how_1933,zwicky_masses_1937} inferred a discrepancy between the mass needed to gravitationally bind the Coma Cluster and the mass estimated from its light. Such inferences depended on equilibrium, virialization and the representativeness of observed velocities. Moving from these dynamical estimates to a cosmological density required further mass accounting based on luminosity and population assumptions, as illustrated by \cite{shapiro_density_1971}. I therefore distinguish the measured cluster velocities from the successive inferences to mass and cosmic density.
\end{enumerate}

\noindent \cite{ostriker_size_1974} and \cite{einasto_dynamic_1974} brought these known observations together under a halo hypothesis. They argued that if each galaxy is embedded in a massive halo extending beyond its visible edge, then:

\begin{itemize}
    \item[(i)] galaxy rotation curves can remain approximately flat;
    \item[(ii)] the added mass associated with cluster galaxies can reduce the virial mass discrepancy; and
    \item[(iii)] the collective mass in galaxy halos could contribute substantially to the cosmic mass budget, potentially bringing $\Omega$ closer to unity.
\end{itemize}

\noindent At the start of their paper, \cite{ostriker_size_1974} write that \textit{``observations may be consistent with a Universe which is `just closed' (}$\Omega = 1$\textit{)''}, adding that some held this conclusion for essentially non-experimental reasons \citep[L1]{ostriker_size_1974}. Their mass estimate nonetheless gives $\Omega\approx0.2$, with an uncertainty of at least a factor of three. Their concluding discussion weighs both low- and high-density arguments, including structure-formation considerations favoring $\Omega\approx1$ \citep[L4]{ostriker_size_1974}. \cite{einasto_dynamic_1974} place greater emphasis on reducing the cluster virial mass discrepancy and likewise infer a galactic matter density around $20\%$ of critical density \citep[310]{einasto_dynamic_1974}. The studies thus connect mass estimates across scales while leaving closure unresolved. A theoretical preference for closure helps explain \textit{why} the mass-budget question was worth pursuing for some researchers, but the case for massive halos still required the astronomical reasoning these papers supplied.

The shared halo framework and its associated parameters made two phenomena previously treated as separate observational anomalies informative about one another. Rotation curves could constrain expectations about cluster dynamics, and cluster dynamics could constrain expectations about rotation curves, through a connection that was not justified when the phenomena were treated in isolation. This changed what could be inferred from the already available observations, marking a transition in 1974 from well-established anomalies to empirical evidence for dark matter. The resulting evidential picture bears directly on the total mass budget of the universe, a question many cosmologists treated as urgent in light of background constraints such as $\Omega \approx 1$.

Two philosophical implications follow from this reconfiguration. First, the shared halo framework allows us to assess how combining rotation curves and cluster dynamics affects confirmation. If the relation between these results is more expected under the halo account than under its alternatives, they can supply comparative support through the informational connection described by Myrvold. The connection and the constraints it imposes therefore matter to the explanation of the 1974 transition.

Second, the same reconfiguration helps explain differences in theory assessment. For researchers concerned with the cosmic mass budget, a cross-scale halo account addressed an especially salient question. Others placed greater weight on unresolved dynamical and observational auxiliaries. The next sections examine how these differences can be represented within a Bayesian assessment, and when the proposed connection contributes comparative support in Myrvold's sense.

\section{Permissible community disagreement}

\noindent Why did the new synthesis persuade some researchers more than others, when they were assessing substantially the same observational results? Their different scientific concerns provide part of the answer: the mass-budget problem gave some cosmologists a reason to pursue connections that researchers working on local dynamics regarded as too dependent on auxiliary assumptions. To assess whether the disagreement was \textit{rationally permissible}, we need to examine the grounds for the background constraints and their role in the inference to missing mass.

\subsection{Evidence and background}

\noindent Rubin’s retrospective makes clear that many astronomers initially resisted treating extensive halos as the default conclusion from rotation-curve data:

\begin{quote}
In 1977, many astronomers hoped that dark matter might be avoided [$\dots$] there were still non-believers. One eminent astronomer said to me,`When you observe low luminosity galaxies, you'll find Keplerian falling rotation curves.' \citep[6]{rubin_brief_2004}
\end{quote}

\noindent She also recalls:

\begin{quote}
Kalnajs’ (1983) insistence that dark matter is not required, at least for a few galaxies with spatially limited data, convinced a few astronomers that dark matter could be avoided. In retrospect, \textit{we think it is fair to say that many astronomers hoped that Kalnajs was right; dark matter was to be avoided, if at all possible.} \citep[7]{rubin_brief_2004} [My emphasis]
\end{quote}

\noindent Rubin's recollections identify a difference in assessment within an overlapping and changing scientific community. Alongside any sociological explanation, we can ask which background assumptions and modeling commitments made the inference to halos more compelling for some researchers than for others.

For researchers concentrating on galactic dynamics, the salient question was whether the data, together with accepted dynamical and astrophysical auxiliaries, justified a robust inference to a new mass component. Given uncertainties about systematics, tracer populations, and particular dynamical models, it was appropriate to explore whether revised auxiliaries could accommodate the anomalies without pervasive unseen matter. Researchers concerned with global cosmological models also asked what the inferred masses implied for the universe's total density. For some, a near-critical density or closure preference made that question especially pressing and encouraged cross-domain arguments. The rational permissibility of these different assessments depends on how well the relevant constraints and auxiliary judgments can be defended.

\subsection{Cosmologists’ priors}

\noindent A Bayesian reconstruction makes the dependence on background assumptions explicit. Let $h$ denote a halo-based dark matter hypothesis (for example, that galaxies are embedded in massive halos), $e$ the relevant empirical results (rotation curves and cluster dynamics), and $b$ the background assumptions fixing modeling commitments and auxiliaries. For each assessment, we specify a probability function and background, giving $\Pr(h \mid e, b)$. Bayes' theorem gives:

\begin{align}\label{Bayes}
\begin{split}
\Pr(h \mid e, b)
&= \frac{\Pr(e \mid h, b)\,\Pr(h \mid b)}{\Pr(e \mid b)} \\
&= \frac{\Pr(e \mid h, b)\,\Pr(h \mid b)}{\Pr(h \mid b)\,\Pr(e \mid h, b) + \Pr(\neg h \mid b)\,\Pr(e \mid \neg h, b)} \,
\end{split}
\end{align}

\noindent This representation distinguishes two ways researchers can differ in their assessments while remaining epistemically responsible.

\begin{enumerate}
    \item \textit{Differences in priors, relative to background assumptions}.\\
Communities can assign different $\Pr(h \mid b)$ because their background constraints make different hypotheses antecedently plausible. In the present case, closure-related considerations and associated cosmological principles could raise the prior probability of additional gravitating mass. These considerations enter through $b$, shaping which hypotheses are treated as serious candidates within the relevant modeling context.

    \item \textit{Differences in likelihoods, due to auxiliaries}.\\
Even when researchers agree on the observational data, they may reasonably differ about $\Pr(e \mid h, b)$ and $\Pr(e \mid \neg h, b)$, since these likelihoods depend on auxiliary assumptions about, for example, equilibrium, tracer populations, mass-to-light ratios, or the reliability of particular measurements. In the 1970s, disagreement about such auxiliaries could make it reasonable for some astronomers to treat the support for halos as weaker or more conditional than cosmologists did when working within a mass-budget framework.
\end{enumerate}

\noindent For a simplified comparison, suppose that, relative to a broadly shared background $b$ concerning mechanics and measurement practices, both communities agree that $h$ makes $e$ comparatively expected. That is, $\Pr(e \mid h, b)$ is high while $\Pr(e \mid \neg h, b)$ is comparatively low. The evidential force of $e$ for $h$ is captured by the Bayes factor:

\begin{align}
\mathrm B = \frac{\Pr(e \mid h, b)}{\;\,\Pr(e \mid \neg h, b)}
\end{align}

\noindent When $B$ is large, $e$ strongly favors $h$ over its negation. In that setting, if the communities’ likelihood assessments are broadly aligned, divergence in $\Pr(h \mid e, b)$ will primarily reflect differences in the prior term $\Pr(h \mid b)$. Historically, closure considerations plausibly affected some prior assessments by making additional gravitating mass antecedently plausible and cross-scale mass accounting worth pursuing \citep{de_swart_closing_2020}. Those considerations operated alongside observational arguments about cosmic density, as the discussion in \cite{ostriker_size_1974} illustrates. Their relative influence must be understood in the context of each assessment.

How far does this justify permissiveness? Belonging to a successful relativistic framework gives a cosmological commitment a context for assessment, but its justification still requires independent motivation, an empirical basis for integration, and openness to revision, as specified in the overview. A commitment meeting these criteria can reasonably shape a prior assessment. The specific evidence for halos must come from comparing the dynamical results under the halo account and its alternatives. We can now ask how much the proposed unification contributes to this comparative support.

\section{Unification}

\noindent The halo synthesis offered more than a reason to pursue the missing-mass question. It connected the results that would bear on the answer. This is the unificatory contribution attributed to the papers by \cite{ostriker_size_1974} and \cite{einasto_dynamic_1974}:

\begin{quote}
The 1974 papers synthesized the two instances of curious galaxy behaviour into a single framework, thereby coalescing the problems into a single anomaly of missing mass. \citep[6]{de_swart_how_2017}
\end{quote}

\noindent The epistemic force attributed to this connection is evident in the description of the studies as ``groundbreaking papers, each presenting a strong case for the existence of large amounts of mass in the outer parts of galaxies'' \cite[21]{bertone_history_2018}.\footnote{Also, in \citealp[L111]{Rubin_1978}: ``The major result of this work is the observation that rotation curves of high-luminosity spiral galaxies are flat, at nuclear distances as great as $r = 50 kpc$. [$\dots$] These results take on added importance in conjunction with the suggestion of Einasto, Kaasik, and Saar (1974), and Ostriker, Peebles, and Yahil (1974) that galaxies contain massive halos extending to large $r$. Such models imply that the galaxy mass increases significantly with increasing $r$ which in turn requires that rotational velocities remain high for large $r$. The observations presented here are thus a necessary but not sufficient condition for massive halos.''} Rubin's retrospective remark identifies what made the connection valuable:

\begin{quote}
Science often advances when ideas, formerly very disparate, are united. In retrospect, it took a long time for astronomers to relate Zwicky’s dark matter to the flat rotation curves for some galaxies that were beginning to attract attention. \citep[3]{rubin_brief_2004}
\end{quote}

\noindent Tremaine similarly characterized the unification as a watershed in galaxy dynamics and cosmology \citep[1223]{tremaine_comments_1999}. If the unification of flat rotation curves and the cluster mass discrepancy by dark matter in 1974 was epistemically significant, we need to explain what the connection contributed.

Rubin's remark locates the epistemic value of unification in a connection between known phenomena. An explanationist can appeal to the virtue of explaining those phenomena through a common origin. For a Bayesian, the difficulty is to identify how the connection affects confirmation when the observations are already known and even helped motivate the hypothesis. Myrvold's account of mutual-information unification allows us to examine the probabilistic contribution first, before returning to the historical novelty of the connection.

\subsection{Unification, explanation, and the Bayesian challenge}

\noindent Much of the philosophical literature treats unification as an \textit{explanatory virtue}, linking it to accounts of scientific inference such as inference to the best explanation (IBE). The guiding thought is that if two theories accommodate the same total evidence, but one provides a more integrated explanation, for example by positing a single cause for otherwise disparate phenomena, we are justified in believing that it is true. This explanatory view is defended as epistemically significant by, among others, \cite{kitcher_explanatory_1989}, \cite{lipton_inference_2003}, and \citet{psillos_scientific_1999, psillos_knowing_2009}. On this construal, the virtue is non-empirical. It concerns the organization and economy of explanation rather than the introduction of new empirical content.

Fitting this idea into Bayesian confirmation theory requires specifying its probabilistic consequences. In Bayesianism, confirmational relations depend on, and are exhausted by, priors and likelihoods relative to background assumptions.\footnote{For treatments of probabilistic reasoning and explanation, see \cite{henderson2014bayesianism}, \cite{weisberg2009locating}, \cite{iranzo2008bayesianism}, and \cite{douven2015probabilistic}.} If two hypotheses assign the same joint likelihood to the evidence under the same background assumptions, unification supplies no additional likelihood-based advantage. This is the challenge for a Bayesian account \citep{howson_scientific_2006,hartmann_bayesian_2011}.

In the dark matter case, the question is which constraints a common halo explanation imposes across the two domains. By examining how those constraints affect the joint likelihood of the established results, we can assess the unification's comparative contribution to confirmation.

\subsection{Mutual-information unification}\label{sec:miu}

\noindent A hypothesis unifies in Myrvold's sense when it makes distinct evidential claims informative about one another \citep{myrvold_bayesian_2003, myrvold_evidential_2017}. A common-origin explanation can do this if its consequences for one result constrain what we should expect of another. The relevant probabilistic property is the dependence introduced between the results by the hypothesis. Let $\mathbf{e} = \{p_1,p_2,\dots,p_n\}$ be a body of evidential propositions. Following \citet[eq.~(7)]{myrvold_bayesian_2003}, we define their informational relevance by

\begin{equation}\label{eq:pointwise-information}
I(p_1;p_2\mid b)
=\log\frac{\Pr(p_1\land p_2\mid b)}{\Pr(p_1\mid b)\Pr(p_2\mid b)}
\end{equation}

\noindent The quantity measures informational relevance point-wise, for the particular propositions under assessment. Unlike expected Shannon mutual information, it can be negative: its sign tells us whether one proposition makes the other more probable, leaves its probability unchanged, or makes it less probable. All ratios below require positive probabilities. If $I(p_1;p_2\mid b)\approx0$ but $I(p_1,p_2\mid h,b)>0$, the hypothesis makes the two results positively relevant to one another. Myrvold captures the unificatory contribution by the following definition:

\begin{align}
U(p_1,p_2;h \mid b)\;=\; I(p_1;p_2 \mid h \land b) \;-\; I(p_1;p_2 \mid b) \,
\end{align}

\noindent In the idealized case where $I(p_1;p_2 \mid b)=0$, $U$ equals $I(p_1;p_2 \mid h \land b)$. A positive value identifies an evidential connection absent from the background assessment. To connect this to confirmation, it is useful to work with an information-theoretic measure of evidential support, equivalently a log Bayes factor:

\begin{align}
C(h; e \mid b)= \log \frac{\Pr(e \mid h, b)}{\;\;\ \Pr(e \mid \neg h, b)} \,
\end{align}

\noindent For this log-Bayes-factor measure, Myrvold's decomposition compares the unification provided by $h$ with that provided by its alternative \citep[eq.~(13)]{myrvold_bayesian_2003}:

\begin{align}
C(h; e_1 \land e_2 \mid b)
\;=\;
C(h; e_1 \mid b)
\;+\;
C(h; e_2 \mid b)
\;+\;
U(e_1,e_2; h \mid b)
-U(e_1,e_2;\neg h\mid b) \,
\label{eq:comparative-unification}
\end{align}

\noindent Here the difference between the two unification terms determines the additional comparative support. Even a positive $U$ for $h$ supplies no additional unificatory advantage if the alternative generates an equally strong connection. Myrvold's simpler decomposition with just $U(h)$ applies to the different measure based on the log posterior-to-prior ratio \citep[eq.~(12)]{myrvold_bayesian_2003}. For the Bayes-factor comparison used here, we must specify the alternatives and their weights to evaluate $\Pr(e\mid\neg h,b)$. The question is then whether the halo framework's particular constraints make the established combination more expected than the competing accounts do.

There are reasons to look beyond informational dependence when assessing unification's epistemic value. \citet{Cast_radin_dawid} identify a further possible source of confirmation in the surprising discovery that a unified account exists at all. Evaluating that contribution requires expectations about theory space and the availability of alternatives, extending the assessment beyond predictions within a theory's intended domain. Applying their argument to 1974 would require contemporary grounds for the relevant expectations and surprise. My concern here is the contribution we can investigate through the halo proposal's cross-domain constraints. Even within that assessment, a likelihood calculation leaves us with the question of whether the connecting auxiliaries are scientifically justified.

\subsection{Evidential reconfiguration in 1974}

\noindent The question now is whether the 1974 synthesis fits this MIU structure. Let $h$ be the halo-based dark matter hypothesis, and let $p_1$ and $p_2$ denote the two relevant phenomena:

\begin{align}
h &\colon \text{galaxies are embedded in extended halos with substantial unseen mass,}\notag \\
p_1 &\colon \text{the galaxy cluster mass discrepancy (virial mass deficit),}\notag \\
p_2 &\colon \text{flat galaxy rotation curves at large radii,}\notag \\
b &\colon \text{background assumptions (GR, ordinary matter, measurement conventions),}\notag \\
\uptheta &\colon \text{halo-to-luminous mass normalization parameter.}\notag
\end{align}

\noindent In a concrete halo model, the parameter space $\Theta$ includes profile shapes, scale radii, truncations and population differences. Here $\uptheta$ denotes a schematic halo mass-to-light normalization. The historical estimates indicate the scale and kind of mass accounting involved: \cite{ostriker_size_1974} estimate a mass-to-light ratio of order $200$ in solar units for giant spirals, while \cite{einasto_dynamic_1974} report values around $100$ for spirals and above $120$ for ellipticals \citep[310]{einasto_dynamic_1974}.\footnote{Each estimate belongs to its analysis, with cross-domain implications that depend on the profile, radial range, and population assumptions retained there. Treating these values as estimates of a single universal parameter would conceal those differences.}

For the shared halo framework to make the results informative about one another, conditioning on one result must revise the distribution over halo parameters in a way that \textit{changes expectations about the other}. The evidential connection therefore depends on how this revision affects the model's predictions.\footnote{Reweighting the parameter values can be sufficient, even if all remain in the posterior support.} The historical contrast begins with the local contexts in which the two phenomena were commonly assessed before the synthesis:

\begin{quote}
Clearly, in neither branch of astronomy were these observations at this point cited as positive evidence for the presence of extra matter, or falsifying evidence for any alternative hypothesis. Furthermore $[\dots]$ the two problems were studied separately. \citep[3]{de_swart_how_2017}
\end{quote}

\noindent Figure \ref{fig:dag-side-by-side} distinguishes between those local assessments (a) and an integrated halo model (b). In the latter, probabilistic relevance is reciprocal: whenever conditioning on one result changes the probability of the other, the reverse assessment also produces a change. To calculate the contribution of the connection, I introduce an uncoupled control in which the two probabilities factorize. Probabilistic independence is an assumption of this control, requiring more than the historical fact that the investigations were separate. In the shared halo model, both results depend on $\theta$. If conditioning on one result reweights the parameter values so that the other result becomes more probable, the results are positively probabilistically relevant under the halo hypothesis. Relative to a background in which their informational relevance is zero, this gives:

\begin{align}
I(p_1;p_2 \mid h \land b) > 0
\quad \text{and therefore} \quad
U(p_1,p_2;h \mid b) > 0 \,
\end{align}

\begin{figure}[htbp]
  \centering

  \begin{subfigure}[t]{0.40\textwidth}
  \centering
  \begin{tikzpicture}
  \node[dag node, minimum width=56.0803811pt] (theta1) at (-2.1,2.01) 
  {$\uptheta_1$: local\\cluster\\parameter};
  \node[dag node, minimum width=56.0803809pt] (p1)     at (-2.1,-1.22) 
  {p1: cluster\\mass\\discrepancy};
  \draw[dag edge, semithick, arrows=-Latex] (theta1) -- (p1);
  \node[dag box, fit=(theta1)(p1), rounded corners=31.2pt] (ctx1) {};
  \node[dag label, anchor=south, yshift=2pt] at (ctx1.north) {Cluster\\context};

  \node[dag node, minimum width=57.7875465pt] (theta2) at (1.2,2.04)
  {$\uptheta_2$: local\\galaxy\\parameter};
  \node[dag node, minimum width=56.6494356pt] (p2)     at (1.22,-1.22) 
  {p2: flat\\rotation\\curves};
  \draw[dag edge, semithick, arrows=-Latex] (theta2) -- (p2);
  \node[dag box, fit=(theta2)(p2), rounded corners=31.2pt] (ctx2) {};
  \node[dag label, anchor=south, yshift=2pt] at (ctx2.north) {Galaxy\\context};
  \end{tikzpicture}
  
  \subcaption{Separate local assessments.}
  \label{dag-a}
  \end{subfigure}
  \hspace{0.1cm}
  \begin{subfigure}[t]{0.5\textwidth}
    \begin{tikzpicture}
    
\node[dag node, minimum width=56.9055122pt] (h)     at (0.11,1.5)          
	{$h$: halo\\framework};
\node[dag node, minimum width=56.3364575pt] (theta) at (0.12,-1.25)  
	{$\uptheta$: halo\\normalization};
\node[dag node, minimum width=58.0436213pt] (p1)    at (-2.52,-1.28) 
	{$p_1$: cluster\\mass\\discrepancy};
\node[dag node, minimum width=58.0436217pt] (p2)    at ( 2.77,-1.28) 
	{$p_2$: flat\\rotation\\curves};
	
\node[dag box, fit=(theta)(p1), 
	rounded corners=35pt, minimum width=70pt, shape=rectangle, minimum height=73.5pt, inner sep=6.5pt, densely dashed, label={[fill=black!15, draw=black, densely dashed, label distance=-3.0pt]255.3:\textbf{A}}] (ctx0) {};
	
\node[dag box, fit=(theta)(p2), rounded corners=37.5pt, minimum width=75.5pt,
	shape=rectangle, minimum height=73.5pt, inner sep=6.5pt, solid, 		label={[fill=black!15, draw=black, solid, label distance=-3.15pt]289.6:\textbf{B}}, semithick] (ctx1) {};
	
\node[dag node, shape=rectangle, fill=black!12, rounded corners=7.9pt, draw=none, text=black, fill opacity=1, minimum height=22pt, minimum width=110pt, text width=3.5cm] (p3) at (-1.85,-4.0) {$\Delta(p_1\to p_2)$: Change in predictive probability for $p_2$};

\node[dag node, shape=rectangle, fill=black!11, rounded corners=7.9pt, draw=none, 
	fill opacity=1, minimum height=22pt, minimum width=110pt, text width=3.5cm] 
	(p4) at (2.18,-4.0) {$\Delta(p_2\to p_1)$: Change in predictive probability for $p_1$};
	
\draw[dag edge, arrows=-Latex, semithick] (h) -- (theta);
\draw[semithick, arrows=-Latex] (theta) -- (p2);
\draw[semithick, arrows=-Latex] (theta) -- (p1);

\draw[arrows=Latex-, semithick, densely dotted] (p3.north) -- (-1.85,-3.0);
\draw[arrows=Latex-, semithick, densely dotted] (p4.north) -- (2.18,-3.0);
\end{tikzpicture}
    \subcaption{Reciprocal inference through a shared halo normalization.}
    \label{dag-b}
  \end{subfigure}

  \caption{Separate and integrated assessments of the same empirical results. Solid arrows 	represent modeling dependencies. In (b), A and B show the two paths where conditioning on one result constrains expectations for the other, mediated by tuning the distribution over the shared halo normalization $\uptheta$. Here $\Delta_{i\to j} =\Pr(p_j\mid p_i,h,b)-\Pr(p_j\mid h,b)$, for $(i,j)=(1,2)$ or $(2,1)$. The sign and magnitude of these changes depend on the likelihoods and parameter distribution.}
  \label{fig:dag-side-by-side}
\end{figure}

\noindent Under a halo framework, flat rotation curves constrain the magnitude and distribution of additional gravitating mass associated with galaxies. Because those same halo parameters determine how much mass galaxies contribute to clusters, the rotation-curve result also alters predictions about cluster dynamics. The inference can run in the other direction as well: cluster mass discrepancies constrain the typical magnitude of galactic halos and thereby predictions about rotation curves. Whether this two-way connection increases support depends on how the observed combination compares with the model's expectations. We can display the mechanism through the posterior predictive distribution:

\begin{align}\label{eq:theta-marginal}
\Pr(p_2 \mid p_1,h,b)
\;=\;
\int_{\Theta}
\Pr(p_2 \mid \uptheta,h,b)\,
\pi(\uptheta \mid p_1,h,b)\,
d\uptheta \,
\end{align}

\noindent Here $\pi$ is a proper parameter density, and the equation assumes conditional independence of the two results given $\uptheta,h,b$. The posterior weights the predictions for $p_2$ at each parameter value. For $p_1$ to change the prediction for $p_2$, it must reweight parameter values at which those predictions differ. The following example shows how parameter learning can produce this predictive change.

\subsection{A toy model comparison}\label{sec:toy}

Suppose a deliberately simplified halo model $M_1$ admits two mass-to-light normalizations, $\theta_L$ and $\theta_H$, with equal prior probabilities. The events $p_1$ and $p_2$ represent the cluster mass discrepancy and the outer rotation-curve result, respectively. Conditional on the normalization, their probabilities are:

\begin{center}
\begin{tabular}{lcc}
\toprule
 & $\theta_L$ & $\theta_H$\\
\midrule
$\Pr(p_1\mid\theta,M_1,b)$ & $1/5$ & $4/5$\\
$\Pr(p_2\mid\theta,M_1,b)$ & $1/5$ & $4/5$\\
\bottomrule
\end{tabular}
\end{center}

\noindent The two events are assumed to be independent, conditional on $\uptheta$, and each event's complement receives the probability left by its table entry. The numbers define a normalized toy model in which a more massive halo population makes both dynamical results more expected under fixed transfer and measurement assumptions.\footnote{Note that these probabilities are \textit{stipulated} for the illustration, not estimates from the 1974 papers.} Each result has marginal probability $1/2$. Conditioning on $p_1$ changes $\Pr(\theta_H\mid M_1,b)=1/2$ to $\Pr(\theta_H\mid p_1,M_1,b)=4/5$. Consequently,

\begin{equation}
\Pr(p_2\mid p_1,M_1,b)
=\frac{4}{5}\times\frac{4}{5}
+\frac{1}{5}\times\frac{1}{5}
=\frac{17}{25},
\end{equation}

\noindent which exceeds its marginal probability $1/2$. Conditioning on the first result raises the probability of the second because $p_1$ favors the higher normalization, where $p_2$ is also more probable. The joint likelihood is $17/50$. For comparison, let $M_0$ permit independent local normalizations, each with the same prior and response probabilities. It can describe two local mass discrepancies without imposing the cross-domain relation. Both single-result likelihoods remain $1/2$, but the joint likelihood is $1/4$. The combination therefore gives:

\begin{equation}\label{eq:toy-bf}
B_{M_1:M_0}(p_1,p_2\mid b)
=\frac{17/50}{1/4}=\frac{34}{25}>1
\end{equation}

\noindent Although each result has the same marginal likelihood under both models, their combination favors $M_1$ because of the relation its shared parameter imposes. We can see how much depends on that particular relation by changing the response probabilities for $p_2$. Setting both to $1/2$ gives $B=1$, since reweighting the normalizations now leaves the expectation for $p_2$ unchanged. Reversing them gives $B=16/25<1$: the normalization favored by $p_1$ reduces the probability of $p_2$. The shared parameter permits all three outcomes.

How far does this comparison take us in assessing the historical episode? $M_0$ is a specified uncoupled control, whereas the alternatives scientists actually considered involved substantive judgments about visible mass distributions and mass-to-light ratios, tracer and projection assumptions, and whether clusters were bound systems in equilibrium. Those judgments could generate their own relations between results. Comparing these alternatives with a halo account requires specifying their likelihoods and proper parameter priors, as equation~\eqref{eq:comparative-unification} makes explicit. Disk-stability arguments also supplied a separate reason to consider halos \citep[L3]{ostriker_size_1974}. These contemporary considerations define the relevant historical comparison. Later modified-gravity proposals illustrate further alternatives to halo accounts, but were unavailable as competitors in 1974.

The toy model establishes how the proposed mechanism works and when it fails. Applying its numerical result to 1974 would require \textit{actual}, justified historical likelihoods and priors, whose estimation is beyond the scope of this paper. The historical argument is that the halo synthesis made a restrictive cross-domain assessment possible, opening the question of how much support its particular constraints supplied relative to defensible alternatives. The calculation reconstructs that contribution through a static model comparison. Conditioning first on $p_1$ displays its informational relevance to $p_2$, while both results were already known to the scientists whose reasoning we are examining. The historical change lay in constructing and justifying the connection that made this assessment available. We can now ask what kind of novelty that amounts to in the context of old evidence.

\section{Old evidence and structural novelty}

\noindent The old-evidence problem is usually framed as follows \citep{glymour_theory_1981}. If an agent assigns probability $1$ to an evidential proposition $e$, then conditionalizing on $e$ cannot change their credences. How, then, can established facts confirm a hypothesis, especially when they helped motivate it? Bayes' theorem makes the difficulty explicit:

\begin{align}
\Pr(h \mid e, b) = \frac{\Pr(e \mid h, b)\,\Pr(h \mid b)}{\Pr(e \mid b)} \, 
\end{align}

\noindent If $\Pr(e\mid b)=1$, coherence also requires $\Pr(e\mid h,b)=1$ whenever the conditional is defined, so the posterior equals the prior. An agent who already treats $e$ as certain therefore needs something further to learn or revise if their credence in $h$ is to change. In the preceding model comparison, by contrast, $b$ specifies the auxiliaries used to assess the results, leaving their likelihoods open to evaluation under each model. We can therefore calculate the support supplied by known results without yet explaining how scientists came to reassess them.

\citet{eva_hartmann_2020} examine such reassessment through the alternatives to a hypothesis. Scientists can become more confident that a hypothesis supplies the only available adequate explanation of known evidence, and Jeffrey conditionalization on that comparative explanatory claim can then increase confidence in the hypothesis. Their account makes the role of alternatives especially clear: finding an explanation has limited force if its competitors explain the same results equally well. The halo comparison addresses a related question by identifying how a cross-domain connection can make known results comparatively informative. The calculation leaves open the stronger claim that the halo proposal was the only available adequate explanation. Establishing that claim would require the further assessment of alternatives emphasized by Eva and Hartmann.

The historical question is how the accepted results acquired their joint evidential role. The halo synthesis changed which propositions could bear on the same hypothesis and which relations among them scientists had reason to accept. This is the change in \textit{evidential structure} identified by \cite{de_swart_how_2017}:

\begin{quote}
$[\dots]$ representations of the establishment of dark matter in terms of an accumulation of evidence miss an essential part of this history: they overlook the necessary conditions that made this very accumulation possible, an accumulation that, at face value, was substantially, even if partly, a reinterpretation of existing observations. Here, theory, along with an institutional shift and expansion in astronomy, played a substantive role: a role that reminds us that simply asking “what was the evidence for missing matter” misses the point; we need to understand why certain observations were eventually conceived as ‘evidence’ of anything in the first place. \cite[6]{de_swart_how_2017}
\end{quote}

\noindent Counterfactual strategies that remove $e$ from the background can help assess the support it supplies, even though scientists continue to know it. For the historical question raised here, we also need to understand what made the new assessment available. The halo connection provides a candidate answer by changing how the accepted results bear on one another.

\subsection{Structural novelty as evidence}

\noindent What was novel about this connection? Two familiar concepts locate novelty in the history of the evidence relative to a theory. \textit{Temporal novelty} concerns evidence unknown before the theory was formulated, while \textit{use novelty} concerns evidence not used in constructing it. \cite{earman_bayes_1992} gives the following confirmation criteria, crediting \cite{worrall1985scientific, worrall1989fresnel} for the latter:

\begin{quote}
\textbf{Temporal novelty}: Suppose that $T \vDash E$. If $E$ was already known to be true prior to the articulation of $T$, then $E$ does not confirm $T$.

\textbf{Use novelty}: Suppose that $T_{1} \vDash E$. and $T_{2} \vDash E$. If $E$ was used in constructing $T_1$, but not in constructing $T_2$, then $T_2$ receives more support from $E$ than does $T_1$. \citep[114]{earman_bayes_1992}
\end{quote}

\noindent The rotation-curve and cluster-dynamics results were already available and helped motivate the halo proposal, so neither kind of novelty characterizes them. Earman's use-novelty criterion still allows used evidence to supply support, giving more support to the theory constructed independently of it. Knowing when the results were obtained and how they were used leaves unanswered what the new cross-domain connection contributed.

Possessing a result and assessing its evidential role are different matters. The rotation-curve measurements and cluster velocities required interpretation through auxiliaries that remained open to dispute. In particular, the cluster mass discrepancy follows from the velocities together with dynamical assumptions. Counting it as a further independent measurement would count the same empirical input twice. The propositions $p_1,p_2$ summarize the interpreted results under the stated auxiliaries. The halo synthesis gave scientists a reason to connect them through a shared parameterization $\uptheta$, allowing each to constrain the evidential significance of the other. I use \textit{structural novelty} to identify this new connection between established results. The simple model demonstrates the positive-relevance claim required for the proposed explanation:\footnote{I use $\perp$ for (conditional) probabilistic independence. Formally,
$p_1 \perp p_2 \mid X$ iff $\Pr(p_1 \land p_2 \mid X) = \Pr(p_1 \mid X)\,\Pr(p_2 \mid X)$, equivalently $\Pr(p_1 \mid p_2 \land X)=\Pr(p_1 \mid X)$ (when the relevant conditional probabilities are defined). Informally, conditional on $X$, learning $p_1$ does not provide any further information about $p_2$ (and vice versa).}

\begin{align}
S \;\colon\; I(p_1;p_2 \mid h \land b) > 0
\quad\text{and}\quad
p_1 \perp p_2 \mid \uptheta,h,b \,
\label{eq:s}
\end{align}

\noindent In \cref{eq:s}, the shared parameter mediates the connection: the results are independent given $\uptheta$, yet positively relevant when we average over it. As the controls in section~\ref{sec:toy} show, the latter property depends on the response probabilities and parameter weights. $S$ gives a concise description of this relation within the model. We assess its contribution through the joint likelihood of $p_1,p_2$ and comparison with the corresponding likelihoods under alternatives. The empirical input remains the established results, with no additional observation of $S$ entering the calculation.

\subsection{When is structural novelty epistemically warranted?}

\noindent If a parameterization is flexible enough, we may be able to accommodate the observations while learning little from their connection. In the halo case, $p_1$ might change the parameter weights too little to appreciably alter the expectation for $p_2$. The framework might also leave profile shapes, radial extents, or population differences unspecified, allowing a fit to $p_1$ with little consequence for $p_2$. Even a restrictive halo model must be assessed against defensible alternatives, some of which may produce a comparable dependence through different auxiliaries.

The uncoupled model $M_0$ defined in section~\ref{sec:toy} provides a control against which we can isolate the contribution of the halo connection. Its joint likelihood factorizes:

\begin{equation}
\Pr(p_1,p_2\mid b,M_0)=\Pr(p_1\mid b,M_0)\,\Pr(p_2\mid b,M_0)
\label{eq:M0}
\end{equation}

\noindent Let $M_1$ be a halo model linking the two domains through a common parameter. Under the stated conditional-independence assumption, its joint likelihood averages over that parameter:

\begin{equation}
\Pr(p_1,p_2\mid b,M_1)=\int_{\Theta}
\Pr(p_1\mid \uptheta,M_1,b)\,\Pr(p_2\mid \uptheta,M_1,b)\,\pi(\uptheta\mid M_1,b)\,d\uptheta
\label{eq:M1}
\end{equation}

\noindent Their comparison is the Bayes factor

\begin{equation}
\mathrm{B}_{M_1:M_0}(p_1,p_2\mid b)
=\frac{\Pr(p_1,p_2\mid b,M_1)}{\Pr(p_1,p_2\mid b,M_0)}
\label{eq:Bmodel}
\end{equation}

\noindent The finite example in section~\ref{sec:toy} implements this comparison with the observational results and assessment background fixed. Because the models differ in their cross-domain commitments, their joint likelihoods can differ even when their single-result likelihoods agree. As above, $b$ specifies the auxiliaries for evaluating $p_1,p_2$, rather than a scientist's total belief state in which both are already certain. The calculation assesses the connection's contribution to comparative support. Explaining the historical transition also requires establishing why scientists had reason to construct and accept it.

\citet{dardashti_hawking_2019} make the need to justify a mediating connection explicit in their Bayesian analysis of analog experiments. Universality arguments support assumptions linking the empirical adequacy of source and target models, giving scientists grounds to apply a relationship whose mathematical form alone would leave its empirical relevance unsettled. In the dark-matter case, the available astronomical results are connected directly through the halo account, whereas an analog experiment uses a laboratory source to investigate an inaccessible target. In both cases, the inference depends on justified assumptions linking the domains. For the halo synthesis, we must ask why the shared mass accounting was a defensible description of galaxies and clusters.

\subsection{MIU and old evidence}

\noindent We can now locate the contribution to the old-evidence problem more precisely. The halo synthesis made familiar results jointly informative through a cross-domain connection. MIU explains how that connection can contribute to comparative support when it makes the observed combination more expected under the halo account than under its alternatives. This is the probabilistic consequence of the proposed reconfiguration, evaluated once the models and their auxiliaries are specified.

To explain rational credence change between the historical assessments, we would need to identify what scientists learned or revised in accepting the connection and how this affected their beliefs. The present argument supplies one part of that explanation by identifying the new evidential role made available in 1974 and reconstructing the support it could provide. Its answer to the historical puzzle is that the same results became evidence in a new way because scientists could now assess what their relation implied.

\section{Conclusion}

\noindent The 1974 papers by \cite{ostriker_size_1974} and \cite{einasto_dynamic_1974} are recognized as pivotal in the evidential consolidation of the dark matter hypothesis, despite relying on observations already available. I have argued that their epistemic contribution can be understood through the new evidential roles those observations acquired within a shared halo framework.

Within this framework, rotation curves and cluster dynamics could constrain one another. The simple model makes the inference explicit: a result changes the assessment of a common halo parameter, which alters expectations for the other result. Comparison with the uncoupled control then shows how the observed combination can favor the connected model, while the modified response probabilities show when it fails to do so. The historical studies motivate this reconstruction. The toy model's numerical values illustrate the mechanism, leaving the strength of the historical support dependent on the relevant likelihoods, priors, and alternatives.

Researchers' scientific concerns and background commitments help explain differences in their reception of the halo connection. Some gave greater weight to the cosmic mass budget, others to unresolved local auxiliaries, with concerns overlapping across astronomy and cosmology. Their assessments could be rationally permissible insofar as the commitments were independently motivated, empirically fruitful, and open to revision. That judgment rests on the reasons available to them at the time.

The observations were familiar, yet their evidential role changed. \textit{Evidential reconfiguration} identifies this historical transformation, and \textit{structural novelty} locates the new connection through which the results became jointly informative. The Bayesian comparison shows how this connection can contribute to confirmation, while a full account of historical belief change would also have to explain its acquisition and acceptance. The 1974 episode shows the work required for observations to function as evidence: scientists needed a defensible relation between the local discrepancies to assess their joint implications for missing mass.

\bibliography{Library}
\bibliographystyle{apacite}

\end{document}